# ON SPATIAL EXTREMES: WITH APPLICATION TO A RAINFALL PROBLEM

By T. A. Buishand, L. de Haan[1] and C. Zhou

*Royal Netherlands Meteorological Institute* (*KNMI*),
*Erasmus University Rotterdam and University of Lisbon
and Erasmus University Rotterdam and Tinbergen Institute*

We consider daily rainfall observations at 32 stations in the province of North Holland (the Netherlands) during 30 years. Let $T$ be the *total* rainfall in this area on one day. An important question is: what is the amount of rainfall $T$ that is exceeded once in 100 years? This is clearly a problem belonging to extreme value theory. Also, it is a genuinely spatial problem.

Recently, a theory of extremes of continuous stochastic processes has been developed. Using the ideas of that theory and much computer power (simulations), we have been able to come up with a reasonable answer to the question above.

**1. Introduction.** When a damaging flood has occurred extreme rainfall statistics are frequently used to answer questions about the rarity of the event. Engineers often need extreme rainfall statistics for the design of structures for flood protection. A typical question is, for example, what is the amount of rain in a given area on one day that is exceeded once in 100 years? Or, more mathematically, what is the 100-year quantile of the total rainfall in the area on one day? In this paper this question is investigated for a low-lying flat area in the northwest of the Netherlands. The area is shown in Figure 1. Because it roughly covers the province of North Holland, it will shortly be indicated as North Holland.

There are 32 rainfall stations in the area for which daily data were available for the 30-year period 1971–2000. Only the fall season, that is, the months September, October and November, is considered. In this season the likelihood of flooding and its impact are relatively large. Because of the restriction to the fall season, it is reasonable to assume stationarity in time. Stationarity in space, except for location and scale, is also assumed.

Received January 2007; revised January 2008.
[1]Supported in part by the FCT project PTDC/MAT/64924/2006.

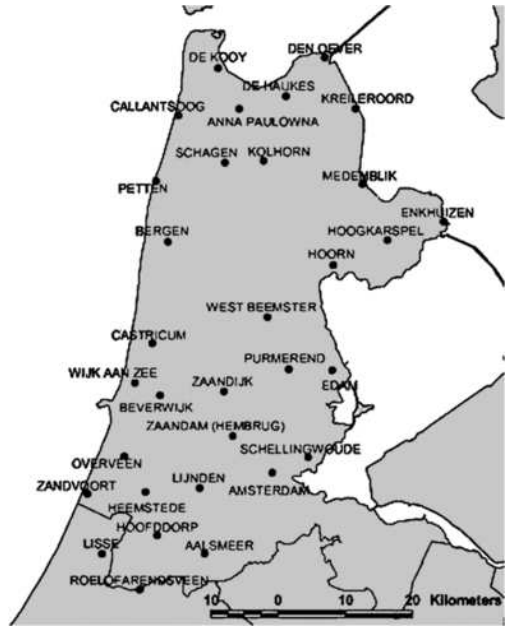

Fig. 1. *The study area: North Holland.*

Since we have to extrapolate from a 30-year to a 100-year period, our problem is an extreme value problem. There is also a clear spatial aspect.

Engineers often make use of areal reduction factors (ARFs) to convert quantiles for point rainfall to the corresponding quantiles of areal rainfall. ARFs have been derived empirically by estimating the areal rainfall as a function of point rainfall measurements [e.g., Natural Environment Research Council (NERC) (1975), Bell (1976)] or by statistical modeling [e.g., Bacchi and Ranzi (1996), Sivapalan and Blöschl (1998), Veneziano and Langousis (2005)]. The latter requires assumptions on distributions, spatial correlation and/or scaling behavior. The resulting ARF for the 100-year quantile is generally very uncertain.

Some attempts have been made to estimate ARFs from weather radar data [Allen and DeGaetano (2005), Stewart (1989)]. One difficulty is that the raw rainfall intensities from the radar reflectivities need to be adjusted for systematic deviations from the values observed at the rainfall stations. Another difficulty is that archived radar data cover a relatively short time period (in the Netherlands only 10 years).

Regional Climate Model (RCM) simulations driven by weather reanalysis data are a potential source for areal aggregated rainfall. A reanalysis is an estimate of the state of the atmosphere based on observations and a numerical weather forecast. The RCM is necessary to increase the spatial resolution.

Various 40-year simulations have been performed recently with spatial resolutions of 50 km × 50 km and 25 km × 25 km, in particular, within the framework of the EU-funded project ENSEMBLES (www.ensembles-eu.org). In addition to the limited length and rather coarse resolution for our application, there are systematic differences between simulated and observed rainfall. For the KNMI RCM driven by ERA40 reanalysis data, Leander and Buishand (2007) report differences up to 20% in seasonal average rainfall for the river Meuse basin, situated south of the Netherlands. For North Holland, the differences can even be larger because it is much smaller than the Meuse basin and it is surrounded by water.

Statisticians have used max-stable processes to obtain the quantiles of the distribution of spatially aggregated rainfall. Coles and Tawn, in a series of papers [Coles (1993), Coles and Tawn (1996)], have developed methods to deal with spatial extremes based on the spectral representation [de Haan (1984), see also Schlather (2002)]. Here we follow a different approach based on random fields. Apart from the parameters that characterize the upper tail of the marginal distributions, our model has a parameter that determines spatial dependence. This parameter is estimated from the tails of the empirical two-dimensional marginal distributions of daily rainfall in North Holland. The 100-year quantile of the total rainfall over the area is found by simulating synthetic daily rainfall fields using the estimated model.

In order to motivate our solution, we first explain some relevant aspects of extreme value theory, in $\mathbb{R}^1$, $\mathbb{R}^d$ ($d > 1$) and $C[0, 1]$ (Section 2). In Section 3 we specify the stochastic process used in the simulation. This process is used only to simulate "extreme" rainfall. For nonextreme rainfall, we sample from the available data. In Section 4 we explain how we combine the two to get a simulated day of rainfall. The estimation of the dependence parameter is dealt with in Section 5. Section 6 discusses the outcome of the simulation and the answer to our problem. Section 7 summarizes our main conclusions.

**2. Extreme value background.** We now explain the background of our approach by reviewing some aspects of extreme value theory and the related theory of excursions over a high threshold. This will be done first in the one-dimensional case (Section 2.1), then the finite-dimensional case (Section 2.2) and finally the case of continuous stochastic processes (Section 2.3). The results in the various cases are similar but of increasing complexity. That is why we start with the one-dimensional case which is well known [Gnedenko (1943) and Pickands (1975), resp.].

2.1. *One-dimensional space.* Suppose that the distribution function $F$ is in the domain of attraction of an extreme value distribution, that is, if $X_1, X_2, \ldots$ are i.i.d. with distribution function $F$, there are a positive function $a$ and a function $b$, such that

$$\lim_{n \to \infty} P\left(\max_{1 \leq i \leq n} \frac{X_i - b(n)}{a(n)} \leq x\right) = G(x),$$

a nondegenerate distribution function. We denote this by $F \in \mathcal{D}$. Then $a$ and $b$ can be chosen such that

$$G(x) = G_\gamma(x) = \exp\{-(1 + \gamma x)^{-1/\gamma}\}$$

for all $x$ with $1 + \gamma x > 0$. Then we also say $F \in \mathcal{D}(G_\gamma)$.

Let $X$ be a random variable with distribution function $F$. Then there exist a positive function $a_0$ and a real shape parameter $\gamma$ (the *extreme value index*), such that for all $x$ with $1 + \gamma x > 0$,

$$\lim_{t \uparrow x^*} P\left(\frac{X - t}{a_0(t)} > x \bigg| X > t\right) = (1 + \gamma x)^{-1/\gamma} =: 1 - Q_\gamma(x).$$

Here $x^* := \sup\{x : F(x) < 1\}$. This means that the larger observations in a sample follow approximately the probability distribution $Q_\gamma$—the generalized Pareto distribution [GPD, cf. Balkema and de Haan (1974), Pickands (1975)]. Note that $1 - Q_\gamma(x) = -\log G_\gamma(x)$.

Let $R$ be a random variable with distribution function $Q_\gamma$. Then,

$$P\left(\frac{R - t}{1 + \gamma t} > x \bigg| R > t\right) = P(R > x)$$

for those $x$ and $t$ for which $1 + \gamma t > 0$ and $1 + \gamma x > 0$. We can call this property *excursion stability*.

Suppose that we have observed a sample $X_1, X_2, \ldots, X_n$ from $F$. Since the approximate distribution of the large values is completely specified, it is possible to use it as a basis to simulate more "large observations," even larger than those in the sample. Thus, by resampling the nonextreme part of the sample and simulating extreme observations from the GPD distribution, one can produce more and more "observations," even extreme ones. Using partly simulation and partly resampling is the main idea behind what we intend to do. Hence, we sample from

(1) $$\bar{F}(x) = \begin{cases} F(x), & \text{if } x < t, \\ 1 - (1 - F(t))Q_\gamma\left(\frac{x - t}{a_0(t)}\right), & \text{if } x \geq t. \end{cases}$$

We can implement this by letting $t$ be one of the upper order statistics and using estimators for $F$, $\gamma$ and $a_0$.

The extra "extreme" observations are sampled from the tail model and they are independent of the "nonextreme" observations. This is justified by the so-called "découpage de Lévy" stating roughly that cutting up a sequence of i.i.d. random variables in two subsequences according to whether their values are in a set $B$ or in its complement $B^c$ results in two independent i.i.d. sequences. The result has been formulated and proved carefully in Resnick (1987), pages 212 and 215. A similar argument applies in higher dimensional space (Sections 2.2 and 2.3).

2.2. *Finite-dimensional space.* Let us now consider the finite-dimensional case, or rather the two-dimensional case, for simplicity. Let $(X, Y)$ be a random vector with distribution function $F$. Suppose $F \in \mathcal{D}$, that is, if $(X_1, Y_1), (X_2, Y_2), \ldots$ are i.i.d. with distribution function $F$, there are positive functions $a$ and $c$ and functions $b$ and $d$, such that

$$\lim_{n \to \infty} P\bigg(\max_{1 \leq i \leq n} \frac{X_i - b(n)}{a(n)} \leq x, \max_{1 \leq i \leq n} \frac{Y_i - d(n)}{c(n)} \leq y\bigg) = G(x, y),$$

a distribution function with nondegenerate marginals. If this is the case, we say $F \in \mathcal{D}(G)$ and $G$ is a (multivariate) extreme value distribution. Then, as in the one-dimensional case, there exists a related two-dimensional GPD distribution $Q_H$, obtained, for example, as follows:

$$\lim_{t \to \infty} P\bigg(\frac{X - b(t)}{a(t)} > \frac{x^{\gamma_1} - 1}{\gamma_1}$$
$$\text{or } \frac{Y - d(t)}{c(t)} > \frac{y^{\gamma_2} - 1}{\gamma_2} \bigg| X > b(t) \text{ or } Y > d(t)\bigg)$$
$$= 2 \int_0^1 \max\bigg(\frac{s}{x}, \frac{1-s}{y}\bigg) H(ds) =: 1 - Q_H(x, y),$$

for $(x, y) \in D_H = \{(x, y) : 2 \int_0^1 \max(\frac{s}{x}, \frac{1-s}{y}) H(ds) \leq 1\} \supset \{(x, y) : x, y \geq 2\}$, where $\gamma_1$ and $\gamma_2$ are the marginal extreme value indices, and $H$ is a probability distribution function on $[0, 1]$ with mean $1/2$. Any distribution $H$ with mean $1/2$ may occur [cf., e.g., de Haan and Ferreira (2006), Chapter 6]. $H$ characterizes the dependence in the tail. It is different from the traditional dependence measure: correlation coefficient which measures the dependence at moderate level. If $H$ concentrates all its measure on point $1/2$, $(X, Y)$ is completely tail dependent. If $H$ is a discrete measure on only two points, 0 and 1 with weight $1/2$ each, $(X, Y)$ is completely tail independent. Similar to the one-dimensional case, we have

$$-\log G\bigg(\frac{x^{\gamma_1} - 1}{\gamma_1}, \frac{y^{\gamma_2} - 1}{\gamma_2}\bigg) = 1 - Q_H(x, y).$$

$Q_H$ is a probability distribution function on $D_H$ with the following properties:

1. Standard one-dimensional GPD marginals: $Q_H(x, \infty) = Q_H(\infty, x) = 1 - 1/x$, for $x \geq 1$.

2. Homogeneity: $1 - Q_H(tx, ty) = t^{-1}(1 - Q_H(x,y))$ for $t > 1$ and $(x,y) \in D_H$, in particular, $Q_H \in \mathcal{D}$:

$$Q_H^n(nx, ny) = (1 - (1 - Q_H(x,y))/n)^n$$
$$\to \exp\{-(1 - Q_H(x,y))\} = G\left(\frac{x^{\gamma_1} - 1}{\gamma_1}, \frac{y^{\gamma_2} - 1}{\gamma_2}\right).$$

Sometimes the function $1 - Q_H(1/x, 1/y)$ is called the *asymptotic dependence function* of $F$. It determines the tail dependence between the two components without specifying the marginal distributions.

3. Excursion stability: If $(R, S)$ is a random vector with distribution function $Q_H$, then with $c := 1 - Q_H(1,1)$, we have for $x, y \in D_H$, $t > c$

$$P\left(R > \frac{tx}{c} \text{ or } S > \frac{ty}{c} \bigg| R > t \text{ or } S > t\right) = P(R > x \text{ or } S > y).$$

We remark that a random vector with an arbitrary extreme value distribution can be constructed as follows. Let $E_1, E_2, \ldots$ be i.i.d. standard exponential random variables. Let $V$ be a random variable with distribution function $H$ and consider i.i.d. copies $V_1, V_2, \ldots$ of $V$. Let the sequences $\{E_i\}$ and $\{V_i\}$ be independent. Then the random vector

$$\left(\max_{i \geq 1} 2V_i/(E_1 + E_2 + \cdots + E_i), \max_{i \geq 1} 2(1 - V_i)/(E_1 + E_2 + \cdots + E_i)\right)$$

has an extreme value distribution and both marginals have distribution function $\exp(-1/x), x > 0$.

We want to follow the line of reasoning from the one-dimensional situation and propose to use $Q_H$ to simulate more "large observations," to be combined with resampling from the available sample. However, simulation from a multivariate distribution is more complicated than in the one-dimensional case. It is more convenient if we can find a random vector that is easy to simulate and that has the same distribution. Consider the random vector $(2YV, 2Y(1 - V))$ with $Y$ and $V$ independent, $Y$ has distribution function $1 - 1/x$, $x \geq 1$, and $V$ has distribution function $H$. It is easy to check that the distribution function $Q_H^0(x, y)$ of $(2YV, 2Y(1 - V))$ coincides with $Q_H(x, y)$ for $x, y \geq 2$. The fact that the distribution function is not exactly the same is not a problem: we are dealing with an asymptotic property and the important thing is that $Q_H^0$ has the asymptotic dependence function $1 - Q_H(1/x, 1/y)$, that is,

$$\lim_{t \to \infty} P(2YV > tx \text{ or } 2Y(1 - V) > ty | 2YV > t \text{ or } 2Y(1 - V) > t)$$
(2)
$$= 1 - Q_H(x, y)$$

for $x, y > 1$. In fact, any distribution function in the domain of attraction of $G$ would do since the asymptotic dependence structure is the same as for the limiting extreme value distribution.

Now the random vector $(2YV, 2Y(1-V))$ is useful but not flexible enough: the set of conditions $V \in [0,1]$ and $EV = 1/2$ is rather restrictive. Hence, let us consider the random vector $(YA_1, YA_2)$ with $Y$ and the vector $(A_1, A_2)$ independent, $Y$ as before and $A_1$ and $A_2$ positive with $EA_1 = EA_2 = 1$. The distribution function $Q^*$ of $(YA_1, YA_2)$ satisfies the following properties:

$1^*$. $1 - Q^*(x, \infty) = E\min(1, \frac{A_1}{x})$ for $x > 0$, hence, $\lim_{t \to \infty} t(1 - Q^*(tx, \infty)) = 1/x$, similarly for $Q^*(\infty, x)$;

$2^*$. $\lim_{t \to \infty} t(1 - Q^*(tx, ty)) = E\max(\frac{A_1}{x}, \frac{A_2}{y})$ for $x, y > 0$, that is, $Q^* \in \mathcal{D}$;

$3^*$.
$$\lim_{t \to \infty} P(YA_1 > tx/c \text{ or } YA_2 > ty/c | YA_1 > t \text{ or } YA_2 > t) = E\max\left(\frac{A_1}{x}, \frac{A_2}{y}\right)$$

for $x, y > 0$ with $c := E\max(A_1, A_2)$.

We can easily simulate from $Q^*$, but this distribution satisfies only approximately (not exactly) the three properties 1, 2 and 3. Because of property $2^*$ [meaning that the distribution function of $(YA_1, YA_2)$ has the same asymptotic dependence function as the distribution function of $\max(0, 1 - E\max(\frac{A_1}{x}, \frac{A_2}{y}))$, cf. (2)], we can still use $Q^*$ for simulation, albeit with caution.

2.3. *Extremes of continuous stochastic processes.* What do we mean by extremes in $C[0,1]$, the space of continuous functions defined on the unit interval? The setup is as follows. Let $\{X(s)\}_{s \in [0,1]}$ be a stochastic process in $C[0,1]$. Consider independent copies $X_1, X_2, \ldots$ of the process $X$. Compose for each $n$ a continuous stochastic process

$$\left\{\max_{1 \leq i \leq n} X_i(s)\right\}_{s \in [0,1]}.$$

Suppose that for some positive functions $a_s(n)$ and real functions $b_s(n)$, the sequence of processes

$$\left\{\max_{1 \leq i \leq n} \frac{X_i(s) - b_s(n)}{a_s(n)}\right\}_{s \in [0,1]}$$

converges in $C[0,1]$. If this is the case, we say $X \in \mathcal{D}$. Let us call the limiting process $\{U(s)\}_{s \in [0,1]}$. Then we say $X \in \mathcal{D}(U)$. The following proposition is useful for our purposes [de Haan and Lin (2001)].

PROPOSITION 2.1. *$X \in \mathcal{D}$ if and only if the following two statements hold:*

*1. (Uniform convergence of the marginal distributions.) There exists a continuous function $\gamma(s)$ such that, for $x > 0$,*

$$\lim_{n \to \infty} P\left(\max_{1 \leq i \leq n} \frac{X_i(s) - b_s(n)}{a_s(n)} \leq \frac{x^{\gamma(s)} - 1}{\gamma(s)}\right) = \exp\left(-\frac{1}{x}\right),$$

uniformly for $s \in [0,1]$.

2. (*Convergence of the standardized process.*) With $F_s(x) := P(X(s) \leq x)$ for $s \in [0,1]$,

$$\left\{ \max_{1 \leq i \leq n} \frac{1}{n(1 - F_s(X_i(s)))} \right\} \xrightarrow{d} \{\eta(s)\} \qquad (say)$$

in $C[0,1]$. Note that all one-dimensional marginal distributions of the process $1/(1 - F_s(X_i(s)))$ are equal to $1 - 1/x$, $x \geq 1$.

The process $\eta$ satisfies the following: if $\eta_1, \eta_2, \ldots$ are i.i.d. copies of $\eta$, then

$$\frac{1}{n} \max_{1 \leq i \leq n} \eta_i \stackrel{d}{=} \eta,$$

that is, the process is *simple max-stable*. [The word "simple" indicates that all marginal distributions are standard Fréchet distributions, $\exp(-1/x), x > 0$.] Moreover, we have that

$$\{U(s)\} \stackrel{d}{=} \left\{ \frac{(\eta(s))^{\gamma(s)} - 1}{\gamma(s)} \right\}.$$

As a consequence of this proposition, we can study the "simple" process $\eta$ first and go back to $U$ later, in a straightforward way.

Two characterizations of simple max-stable processes are known. One of them can serve our purposes. It is given in the following proposition. The other characterization is discussed at the end of this subsection.

PROPOSITION 2.2 [Schlather (2002), de Haan and Ferreira (2006), Corollary 9.4.5]. *Every simple max-stable process $\eta$ in $C[0,1]$ can be generated in the following way. Let $E_1, E_2, \ldots$ be i.i.d. standard exponential random variables. Further, consider i.i.d. positive stochastic processes $V, V_1, V_2, \ldots$ in $C[0,1]$ with $EV(s) = 1$ for all $s \in [0,1]$ and $E \sup_{0 \leq s \leq 1} V(s) < \infty$. Let the sequences $\{E_i\}$ and $\{V_i\}$ be independent. Then*

$$\eta \stackrel{d}{=} \max_{i \geq 1} V_i / (E_1 + E_2 + \cdots + E_i).$$

*Conversely, each process with this representation is simple max-stable. One can take the stochastic process $V$ such that*

$$\sup_{0 \leq s \leq 1} V(s) = c \qquad a.s.$$

*with $c$ some positive nonrandom constant.*

Now recall the "generalized Pareto" results in one- and finite-dimensional extremes that allowed us to simulate from the tail of the distribution. What is the situation in this spatial setup?

One way to proceed is as in the finite-dimensional case. Let $Y$ be a random variable with distribution function $1 - 1/x$, $x \geq 1$ (i.e., one-dimensional GPD). Let $V$ be a positive stochastic process in $C[0,1]$ that satisfies the conditions of Proposition 2.2: $EV(s) = 1$ for $s \in [0,1]$ and $\sup_{0 \leq s \leq 1} V(s) = c$, a nonrandom constant. Let $Y$ and $V$ be independent. Consider *the GPD process*

$$\{\xi(s)\}_{s \in [0,1]} := \{YV(s)\}_{s \in [0,1]}.$$

The process $\xi$ is in $C[0,1]$ and satisfies the following:
1. Standard GPD tail: $P(YV(s) > x) = 1/x$ for $x > c$;
2. Homogeneity;
3. Excursion stability: The distribution of $\{cYV(s)/t\}$ given $\sup_{0 \leq s \leq 1} YV(s) > t$ is the same as that of $\{YV(s)\}$ for $t > c$.

The validity of the three properties requires the condition that $\sup_{0 \leq s \leq 1} V(s) = c$, a nonrandom constant. If we only know $E \sup_{0 \leq s \leq 1} V(s) < \infty$, the properties do not hold as they stand, but we still have an asymptotic version of them as in the finite-dimensional case. In particular, $\xi \in \mathcal{D}(\eta)$.

We remark that the stochastic process $\{YV(s)\}$ is in the domain of attraction of the process $\{\eta(s)\}$, hence, the asymptotic dependence structure of the two processes is the same (cf. Section 2.2) and either of the processes can be used for simulating extreme events. This is also true for the process $\{YV(s)\}$ with the weaker side condition $E \sup_{0 \leq s \leq 1} V(s) < \infty$. Hence, there are three candidate processes for simulating extremal rainfall.

We finish this section with two remarks.

In all of the above we can replace $[0,1]$ by any compact subset of an Euclidean space, that is, we can deal with spatial extremes.

An alternative approach to areal rainfall is presented in Coles (1993) and Coles and Tawn (1996), as mentioned before. Rather than the representation of Proposition 2.2, their approach is based on an alternative, more analytical, representation involving spectral functions originating from de Haan (1984) and an unpublished manuscript by Smith (1990). They sketch how to calculate (rather than simulate) a quantile of the areal rainfall. The model developed by Coles (1993) consists of a multivariate extreme-value distribution that describes the extremes at a subset of the rainfall stations and deterministic "storm profile functions" to obtain the amount of rain in the remaining points of the area. A consequence of this approach is that the model depends on the positions of the rainfall stations. Coles and Tawn (1996) applied this model to calculate quantiles of extreme daily areal rainfall in the winter season for a region in south-west England. It is assumed that heavy rainfall always takes place throughout the region. Schlather (2002) advocated the use of the representation of Proposition 2.2 to simulate extreme widespread rainfall, like winter rainfall in south-west England.

**3. Stochastic process for simulating "extreme" rainfall.** The starting point for the simulation of the rainfall process is Proposition 2.2, the representation of simple max-stable processes and its counterpart, the excursion stable process $\{YV(s)\}$. Conceptually, as explained in Section 2, the excursion stable process is the right one to use.

However, nonparametric estimation of the characteristics of the process $V$ is presently beyond our reach. Hence, we choose to work with a tractable parametric model for $V$. Unfortunately, the condition $\sup_{0 \leq s \leq 1} V(s) = c$, that makes the process $\{YV(s)\}$ excursion stable, is very stringent and we could not find a reasonable parametric model for such a process. Hence, it seems better to stay with the model $\{YV(s)\}$ but replace the condition $\sup_{0 \leq s \leq 1} V(s) = c$ by $E \sup_{0 \leq s \leq 1} V(s) < \infty$, as allowed by Proposition 2.2. Then the excursion stability is still approximately true, that is, the process has the same asymptotic dependence structure. But we meet another problem. In order to tie the simulated process to the observed nonextreme rainfall, it is imperative that the marginal distributions of the simulated process have a GPD tail [cf. relation (6) below]. As explained in Section 2, this is not correct for $\{YV(s)\}$ with $E \sup_{0 \leq s \leq 1} V(s) < \infty$, worse, the marginal distribution is quite untractable, hence, a transformation to repair this problem seems difficult to find.

Only the third possibility remains: to choose the simple max-stable process from Proposition 2.2 for the simulation. Then the asymptotic dependence structure of the process is the same as that of the corresponding GPD-type process $\{YV(s)\}$ (they are in the same domain of attraction) and the marginal distributions are all the same, hence, they can easily be transformed to the distribution function $1 - 1/x$, $x \geq 1$. The transformation on marginal distributions will not change the asymptotic dependence structure.

This is what we do in the simulation. For $V$, we choose the so-called exponential martingale [cf. Øksendal (1992), Exercise 4.10]. Also, we have to extend the process to a process with a two-dimensional index set. We choose the model

$$(3) \qquad \eta(s_1, s_2) := \max_{i \geq 1} \frac{\exp\{W_{1i}(\beta s_1) + W_{2i}(\beta s_2) - \beta(|s_1| + |s_2|)/2\}}{E_1 + E_2 + \cdots + E_i}$$

for $(s_1, s_2) \in \mathbb{R}^2$ (or rather the area under study, North Holland). Here $\{E_i\}$ is an i.i.d. sequence of standard exponential random variables. The processes $W_{11}, W_{21}, W_{12}, W_{22}, W_{13}, W_{23}, \ldots$ are independent copies of double-sided Brownian motions $W$ defined as follows. Take two independent Brownian motions $B_1$ and $B_2$. Then

$$(4) \qquad W(s) := \begin{cases} B_1(s), & s \geq 0; \\ B_2(-s), & s < 0. \end{cases}$$

The positive constant $\beta$ reflects the amount of spatial dependence at high levels of rainfall: "$\beta$ small" means strong dependence and "$\beta$ large" means weak dependence. The model assumes that the dependence between extreme rainfall at two locations depends only on the distance between the locations as we shall see later on.

The process $\eta$ satisfies the requirements of Proposition 2.2:

$$E \exp\{W_1(\beta s_1) + W_2(\beta s_2) - \beta(|s_1| + |s_2|)/2\} = 1 \quad \text{for } (s_1, s_2) \in \mathbb{R}^2,$$

and

$$E \sup_{\substack{a_1 \leq s_1 \leq b_1 \\ a_2 \leq s_2 \leq b_2}} \exp\{W_1(\beta s_1) + W_2(\beta s_2) - \beta(|s_1| + |s_2|)/2\} < \infty$$

for all $a_1 < b_1, a_2 < b_2$ real.

By Proposition 2.2, the one-dimensional marginal distributions of (3) are all $e^{-1/x}, x > 0$. The two-dimensional marginal distributions are calculated in de Haan and Zhou (2008). They are invariant under a shift. The same holds for the higher-dimensional marginal distributions [the proof is in de Haan and Zhou (2008)]. Hence, the process is shift stationary as it should be for our application.

The choice of this particular process is mainly one of convenience: the process is not too crude and it allows easy simulation and estimation of the dependence parameter.

For the simulation of our process, we need to simulate (3), the maximum of infinitely many terms. However, since the denominators form an increasing sequence, one can approximate the process $\eta$ by taking the maximum of only finitely many terms. In fact, it turns out that even 4 terms are sufficient to get a reasonable result.

We have now a simple max-stable process that can be simulated rather well. But—taking into account our discussion of the finite-dimensional case—in fact, we need a process that has generalized Pareto marginals, not the standard Fréchet extreme value distribution as marginals. Hence, we use the process $\eta$ from (3) but transform the marginal distributions to the generalized Pareto distribution $1 - 1/x$, $x \geq 1$:

$$(5) \qquad \xi(s_1, s_2) := \frac{1}{1 - \exp\{-1/(\eta(s_1, s_2))\}}$$

for $(s_1, s_2)$ in the area.

The last step is a further transformation of the marginal distribution that adapts the process to the local shape ($\gamma$), scale ($a$) and shift ($b$) parameters. These parameters can be estimated from each station separately, using the local sample. However, the resulting estimates may not be accurate enough, due to the small sample size (there is a large number of days with no rain).

To increase precision, it is often assumed in the hydrological and climatological literature that the shape parameter $\gamma$ is constant over the region of interest [e.g., Natural Environment Research Council (NERC) (1975), Alila (1999), Gellens (2002), Fowler and Kilsby (2003)]. A reliable estimate of $\gamma$ is then obtained using all extreme values (usually in the literature this concerns the seasonal or annual maxima) in the region. This can be done by combining the extremes into a single record (the so-called station-year method), by averaging a local estimate of $\gamma$ or a skewness statistic over the region of interest, or by maximizing a log likelihood with a common $\gamma$ and local scale and location parameters. For annual maximum daily rainfall in the Netherlands, Buishand (1991) compared the maximum likelihood approach with the averaging of a local estimate of $\gamma$. Almost the same results were obtained.

Here we use the average of the local estimates of $\gamma$. We found the value $\hat{\gamma} = 0.1082$. This value is comparable with the estimates of the shape parameter found for daily maximum rainfall in the winter half-year (October–March) in the Netherlands [Buishand (1983)] and Belgium [Gellens (2002)]. Of course, our model allows $\gamma$ to vary over the area.

The final transformation results into the process

$$(6) \qquad X(s_1, s_2) := \hat{a}_{(s_1,s_2)}(n/k)\left(\frac{\xi(s_1,s_2)^{\hat{\gamma}_{n,k}} - 1}{\hat{\gamma}_{n,k}}\right) + \hat{b}_{(s_1,s_2)}(n/k).$$

Note that

$$P\left(\frac{X(s_1, s_2) - b_{(s_1,s_2)}(n/k)}{a_{(s_1,s_2)}(n/k)} > x\right)$$
$$= P\left(\frac{\xi(s_1, s_2)^\gamma - 1}{\gamma} > x\right)$$
$$= P\left(\eta(s_1, s_2) > \frac{1}{-\log(1 - (1 + \gamma x)^{-1/\gamma})}\right)$$
$$= (1 + \gamma x)^{-1/\gamma},$$

hence, all marginal distributions are GPD.

The estimation for $\gamma$, $a$ and $b$ (cf. Proposition 2.1) at any location is based on the "extreme" part of the local sample, that is, the upper $k$ order statistics of that sample. In the asymptotic theory, when the sample size $n$ is going to infinity, $k$ will go to infinity: $k = k(n) \to \infty$; but of lower order than $n$: $k(n)/n \to 0, n \to \infty$.

The estimation of the shift $b_{(s_1,s_2)}(n/k)$ is particularly simple: $\hat{b}_{(s_1,s_2)}(n/k)$ is the $k$th largest order statistics of the local sample. There are various estimators of $\gamma$ and $a_{(s_1,s_2)}(n/k)$ that converge at speed $k^{-1/2}$. In the present application we use the so-called moment estimator for $\gamma$ [cf., e.g., de Haan and Ferreira (2006), Section 3.9] and the accompanying estimator for $a_{(s_1,s_2)}(n/k)$ [cf., e.g., de Haan and Ferreira (2006), Section 4.2].

As explained before, to obtain a global shape parameter, we take the average of the local estimates of $\gamma$ among all the stations. However, we keep the local estimates of the scale and shift at each station.

To choose the number of upper order statistics $k$ used for the estimation of the shape parameter, we plot the average estimate of this parameter across monitoring stations against $k$. This average is constant when $k$ is around 125. Similar plots for each individual station confirmed that to choose $k = 125$ is also reasonable for most stations. Therefore, we keep it also for estimating the scale $a$ and the shift $b$ throughout the area. The sample size $n$ is 2730.

With these estimations, the process (6) provides the simulated (extreme) rainfall in the area.

**4. Simulating a day of rainfall.** On an arbitrary day, there will be "extreme" rainfall in part of the area and "nonextreme" rainfall (or no rainfall at all) in the rest of the area.

We achieve this in the simulation as follows: on the one hand, we simulate the process (6) for the whole area; on the other hand, we choose at random a day out of the $30 \times (30 + 31 + 30) = 2730$ days of observed rainfall and we connect the two as follows:

For each station we check whether the observed rainfall on the chosen day is larger than the shift parameter $\hat{b}_{(s_1,s_2)}(n/k)$ for that station. If so, we use (6) (i.e., the simulated process) to get the rainfall at that station. If not, we just use the observed rainfall for the chosen day at that station.

How do we extend this to obtain the rainfall in the entire area?

First we connect the monitoring stations with each other, so as to cover the area with Triangles. The division is presented in Figure 2 as the solid lines. The numbers refer to the local 100-year quantile for each station; see Section 6. We write Triangles since later on we shall also deal with smaller triangles, also we write Vertex and Edge for a vertex and edge of a Triangle. Any Triangle can be extreme or nonextreme.

1. Nonextreme: this is the case if all Vertices of the Triangle are nonextreme. The rainfall in such a Triangle is just a linear function whose values at the Vertices are the observed values.

2. Extreme: all other cases. In that case the rainfall is mainly determined by the process (6) where the functions $a_{(s_1,s_2)}(n,k)$ and $b_{(s_1,s_2)}(n,k)$ on the Triangle are chosen as linear functions whose values at the Vertices are the values obtained by local estimation. Note that this mix of extreme and nonextreme simulation is similar to (1).

More specifically we proceed as follows:

2a. Subdivide each Edge into $d$ intervals of equal length. Connect the separating points on the Edges with each other using lines parallel to the

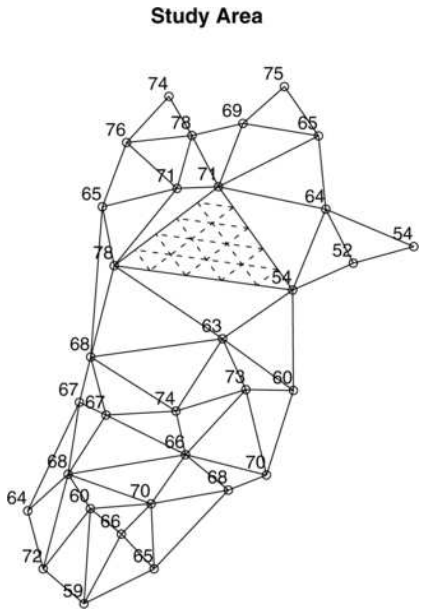

Fig. 2. *The Triangles connecting the observation stations (the numbers give the local 100-year quantile in mm).*

Edges as indicated by the dashed lines in Figure 2. This results into $d^2$ triangles inside a Triangle. We used $d = 5$ in the simulation.

2b. Next we determine the rainfall process in each vertex (i.e., vertex of a triangle). For Vertices we already determined the process. For the vertices, there are two cases.

2b.1. On an Edge connecting two nonextreme Vertices in an extreme Triangle, the rainfall is chosen to be the linear function whose values at the Vertices are the observed values. This determines the rainfall for all vertices on such an Edge. The process (6) plays no role.

2b.2. The rainfall for every other vertex in an extreme Triangle is determined by the process (6).

2c. In order to carry out the numerical integration, we simplify the rainfall process on each triangle in an extreme Triangle. The rainfall in each triangle is given as a linear function whose value at the vertices is the one obtained in part 2b.

This is the way we obtained a day of rainfall. An example of a simulation for Oct 11, 1997, is given in Figure 3. Rainfall for this day is extreme over large parts of the north and the middle of the study area. The left figure is based on the real data for Oct 11, 1997 and the right one presents a simulation. The latter produces very extreme rainfall over a small region in the middle of the study area with a steep gradient to the south. Note that the process is continuous and that it is easy to integrate numerically.

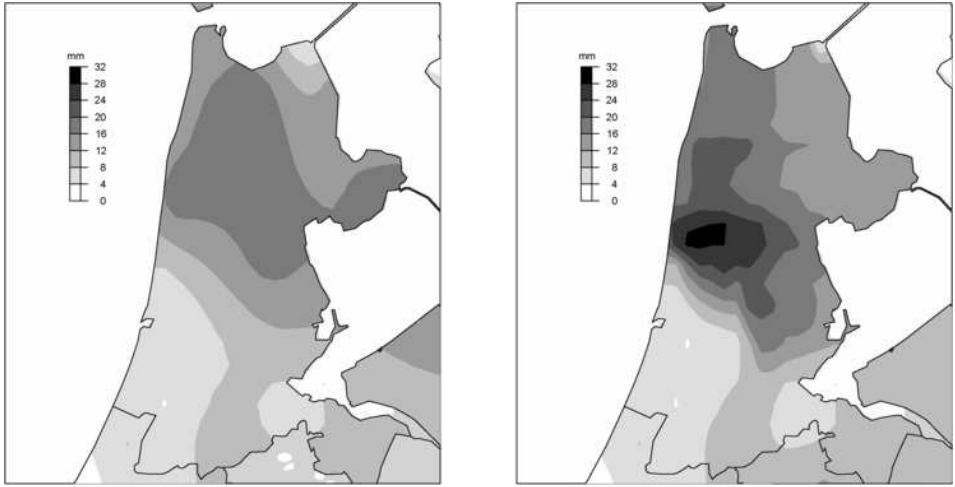

Fig. 3. *Observed (left) and simulated (right) rainfall for Oct 11, 1997.*

We remark that on 2299 out of the 2730 days of observation, none of the Vertices (stations) is extreme, so that no simulation is necessary. On the other hand, there are 44 days on which all Triangles are extreme, so that the whole area is simulated.

The choice for Triangles with monitoring stations as Vertices is one of convenience: triangles fit together easily to produce a continuous process and are relatively simple to handle.

**5. Estimation of the dependence parameter.** One problem remains: we do not know $\beta$, the global dependence parameter in (3). It has to be estimated. This can be done along the lines indicated in de Haan and Pereira (2006).

We need to calculate the two-dimensional marginal distributions of the process $\eta$ [defined in (3)] at locations $(u_1, u_2)$ and $(v_1, v_2)$, say. This is done in de Haan and Zhou (2008). The result is as follows: for $x, y$ real with $h := |u_1 - v_1| + |u_2 - v_2|$,

$$
\begin{aligned}
&P(\eta(u_1, u_2) \leq e^x, \eta(v_1, v_2) \leq e^y) \\
&\quad = \exp\left\{-\left(e^{-x}\Phi\left(\frac{\sqrt{\beta h}}{2} + \frac{y-x}{\sqrt{\beta h}}\right) + e^{-y}\Phi\left(\frac{\sqrt{\beta h}}{2} + \frac{x-y}{\sqrt{\beta h}}\right)\right)\right\},
\end{aligned}
\tag{7}
$$

where $\Phi$ is the standard normal distribution function. Taking $x = y = 0$, we find

$$P(\eta(u_1, u_2) \leq 1, \eta(v_1, v_2) \leq 1) = \exp\left\{-2\Phi\left(\frac{\sqrt{\beta h}}{2}\right)\right\},$$

and consequently,

$$\beta = \frac{4}{h}\left(\Phi^{\leftarrow}\left(-\frac{1}{2}\log P(\eta(u_1,u_2) \leq 1, \eta(v_1,v_2) \leq 1)\right)\right)^2.$$

Hence, we can estimate $\beta$ if we know how to estimate

$$L_{(u_1,u_2),(v_1,v_2)}(1,1) := -\log P(\eta(u_1,u_2) \leq 1, \eta(v_1,v_2) \leq 1).$$

This is a problem of two-dimensional extreme value theory that has been solved by Huang and Mason [cf. Huang (1992), Drees and Huang (1998)].

Let the continuous process $X$ be in $\mathcal{D}$ (cf. beginning of Section 2.3). Let $X_1, X_2, \ldots$ be i.i.d. copies of $X$. Write $\{X_{i,n}(s_1,s_2)\}_{i=1}^n$ for the order statistics at location $(s_1, s_2)$. Then the estimator

$$\hat{L}^{(k)}_{(u_1,u_2),(v_1,v_2)}(1,1) := \frac{1}{k}\sum_{j=1}^n 1_{\{X_j(u_1,u_2) \geq X_{n-k+1,n}(u_1,u_2) \text{ or } X_j(v_1,v_2) \geq X_{n-k+1,n}(v_1,v_2)\}}$$

is consistent provided $k = k(n) \to \infty$, $k(n)/n \to 0$, $n \to \infty$. It is asymptotically normal under certain mild extra conditions.

Now label the monitoring stations with the numbers $1, 2, \ldots, N$ ($N = 32$) and define for $p < q \leq N$,

$$\hat{\beta}_{p,q} = \frac{4}{h}\left(\Phi^{\leftarrow}\left(\frac{1}{2}\hat{L}^{(k(p,q))}_{(u_1,u_2),(v_1,v_2)}(1,1)\right)\right)^2,$$

where $(u_1, u_2)$ and $(v_1, v_2)$ are the coordinates of station $p$ and $q$ respectively, $k(p,q)$ is the number of higher order statistics used in the estimation. Our estimator for $\beta$ is

$$\hat{\beta} := \frac{2}{N(N-1)}\sum_{q=2}^N \sum_{p=1}^{q-1} \hat{\beta}_{p,q}$$

(consistent and asymptotically normal).

We found that $\hat{\beta} = 0.04277$.

Note that the estimators $\hat{\gamma}$, $\hat{a}$ and $\hat{b}$ come from one-dimensional extreme value theory, the estimator $\hat{\beta}$ comes from finite-dimensional extreme value theory and the process $\eta$ comes from extreme value theory in $C[0,1]$.

**6. Application.** Our purpose is to study extremes of the total rainfall in North Holland. In particular, we want to determine how severe the areal rainfall is that occurs once in 100 years. To be precise, it is once in $100 \times (30 + 31 + 30) = 9100$ days. In other words, we are studying the $1 - 1/9100$ quantile of the daily total rainfall in the area. This quantile will be briefly indicated as the 100-year quantile.

Before presenting the simulation result, we would like to introduce some statistics and results for separate stations. Take Station West Beemster as an example (it is located in the middle of the area, and considered as the origin point when simulating the dependence process). The largest observed rainfall in the 30 years is 68.2 mm. By fitting the GPD with shape parameter $\hat{\gamma} = 0.1082$ to the observed extreme daily rainfall amounts at West Beemster, we can estimate the $1 - 1/9100$ quantile for this station. The point estimator is 63.0 mm.

The $1 - 1/9100$ quantiles for the other monitoring stations were obtained in the same way. Figure 2 gives the result for each station. We get that the average $1 - 1/9100$ quantile among all the stations is 66.9 mm.

The simulation procedure in Section 4 has been repeated 91,000 times. This results in a sample of 91,000 days of rainfall in North Holland. For each day we calculate the total rainfall as the numerical integral of the rainfall process on the area. We take the 10th largest order statistic of this sample, that is, we determine the $1 - 1/9100$ sample quantile of the integrated rainfall. Dividing by the total area, 2010 km$^2$, we get the average rainfall in the area. We replicate this procedure 60 times. A histogram of the 60 simulated quantiles is given in Figure 4.

Some statistics of the 60 simulated quantiles are given in Table 1. From the table, the sample mean of the simulated quantiles is 58.8 mm, and the sample standard deviation is 3.16 mm. Hence, the standard deviation of the sample mean is 0.41 mm.

When estimating the dependence parameter $\beta$ in Section 5, we take the average of 496 estimates from all pairs of the monitoring stations, 0.04277.

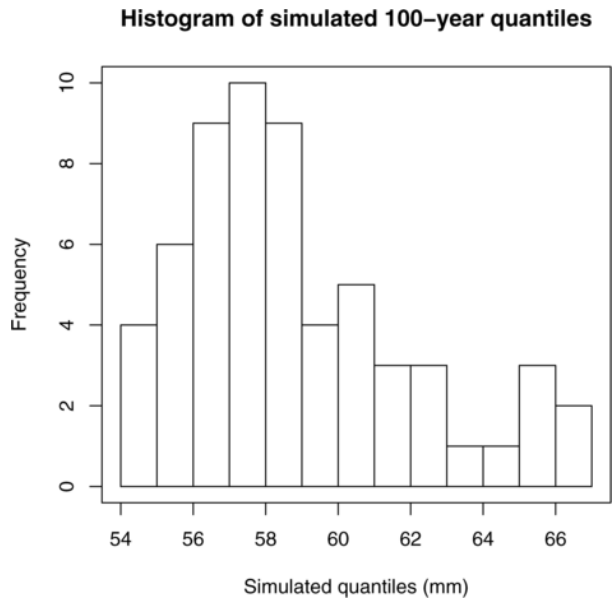

FIG. 4. *Histogram of simulated 100-year quantiles.*

Table 1
*Statistics of simulated 100-year quantiles of area-average rainfall*

| Mean (mm) | 58.8 | Sample Std (mm) | 3.16 |
|---|---|---|---|
| Min | 54.4 | Max | 66.7 |

This is what we used in the above simulation. In order to study the sensitivity of the dependence parameter $\beta$, we take the 25% and 75% quantiles of the 496 estimates, 0.0339 and 0.0496, as $\beta$ in the simulated model to repeat the above analysis. From 10 simulated 100-year quantiles for each new $\beta$, we get sample means of 58.4 mm and 60.0 mm respectively. Hence, the result does not change much.

A setback of the exponential martingale model is the dependence on the coordinate axes. In order to see how important the choice of the axes is, we have repeated the analysis after a 45 degree rotation of the axes and the result does not change a lot (from 10 simulated 100-year quantiles, we get a sample mean of 58.2 mm, and a sample standard deviation of 3.1 mm). After rotation, one of the axes is more or less the prevailing wind direction.

It is interesting to compare the estimated 100-year quantile with the value obtained by fitting the GPD with $\hat{\gamma} = 0.1082$ to the extremes of the average daily rainfall of the 32 stations in the area. The use of the same shape parameter as that for the extremes at the individual stations can be justified from multivariate extreme value theory [Coles and Tawn (1994)]. The resulting value of 57.8 mm for the 100-year quantile is slightly smaller than the average of our simulations.

The quantile for the area-average rainfall is smaller than the average of the corresponding quantile for the individual measurement stations. The areal reduction factor ARF is the ratio of these two quantities, ARF = $58.8/66.9 = 0.88$. It is remarkable that from the graph in the UK Flood Studies Report [see Natural Environment Research Council (NERC) (1975)], a similar value of ARF is found for an area of 2010 $km^2$. The latter refers to annual maximum rainfall rather than seasonal maximum rainfall. The ARF from the GPD fit to the extreme average daily rainfall equals $57.8/66.9 = 0.86$.

**7. Conclusion.** The theory of extremes of continuous processes was used to estimate the 100-year quantile of the daily area-average rainfall over North Holland. The estimation of this quantile was done by simulating the daily process.

Regions with large rainfall were generated using a specific max-stable spatial process. It was argued that direct simulation from the excursion process is not feasible.

The estimated 100-year quantile for the areal average rainfall turns out to be 12% lower than the average 100-year quantile of the 32 measurement stations.

The model used in this paper is convenient: it is really infinite-dimensional (does not depend on a finite number of random variables only); the two-dimensional distributions can be calculated and it can be simulated easily. A disadvantage is that the process is not invariant with respect to rotation of the coordinate axes (see end of Section 6). This will be the subject of future research.

**Acknowledgments.** We are grateful for the thorough comments of three referees, the Associate Editor and the editor. We thank J. Nellestijn for producing Figures 1 and 3.

T. A. BUISHAND  
ROYAL NETHERLANDS METEOROLOGICAL INSTITUTE (KNMI)  
P.O. BOX 201  
3730AE DE BILT  
THE NETHERLANDS  
E-MAIL: Adri.Buishand@knmi.nl  

L. DE HAAN  
DEPARTMENT OF ECONOMICS H08-28  
ERASMUS UNIVERSITY ROTTERDAM  
P.O. BOX 1738  
3062PA ROTTERDAM  
THE NETHERLANDS  
E-MAIL: ldehaan@few.eur.nl  

C. ZHOU  
TINBERGEN INSTITUTE H09-32  
ERASMUS UNIVERSITY ROTTERDAM  
P.O. BOX 1738  
3000DR ROTTERDAM  
THE NETHERLANDS  
E-MAIL: zhou@few.eur.nl